# Complementary-polarity double-layer LiTaO₃ resonators for symmetry-selective SH₂ excitation with ultrahigh electromechanical coupling ($k_t^2$ = 25.7%)

Hao Yan[1†], Zhen-hui Qin[1†], Zhi-Wen Wang[1], Shu-Mao Wu[1], Chen-Bei Hao[1], Hua-Yang Chen[1], Sheng-Nan Liang[1], Ke Chen[4], Si-Yuan Yu[1,2,3]\* and Yan-Feng Chen[1,2,3]\*

[1]National Laboratory of Solid-State Microstructures & Department of Materials Science and Engineering, Nanjing University, Nanjing 210093, China.

[2]Collaborative Innovation Center of Advanced Microstructures, Nanjing University, Nanjing 210093, China

[3]Jiangsu Key Laboratory of Artificial Functional Materials, Nanjing University, Nanjing, 210093, China

[4]School of Electronic Science and Engineering, Nanjing University, Nanjing, 210093, China

† These authors contributed equally: Hao Yan, Zhen-Hui Qin

\*Corresponding authors. e-mail: yusiyuan@nju.edu.cn; yfchen@nju.edu.cn

**Abstract:** We report a structurally simple double-layer lithium tantalate (LiTaO₃) bulk acoustic resonator that enables symmetry-selective excitation of the second-order thickness-shear (SH₂) mode with ultrahigh electromechanical coupling. Two 31° Y-oriented single-crystal LiTaO₃ films are rotation-bonded with complementary polarization (+X/−X) and driven by a longitudinal electric field. Matching between the effective piezoelectric symmetry and the SH₂ mode yields an effective electromechanical coupling coefficient of $k_t^2$ = 25.7% at 5.24 MHz. To our knowledge, this is the highest $k_t^2$ reported for a LiTaO₃ resonator architecture to date. The measured response is dominated by the target SH₂ mode, with only weak parasitic features in the operating band. The structure is also tunable: the resonance frequency and coupling can be adjusted through geometric parameters while maintaining stable modal behavior, indicating good process tolerance. Finite-element analysis further suggests straightforward frequency scaling beyond 5 GHz by reducing the film and electrode thickness while preserving ~25% $k_t^2$. In addition, introducing a SiO₂ compensation layer is predicted to improve the temperature coefficient of frequency to approximately −25 ppm/°C. These results establish complementary-polarity double-layer LiTaO₃ as a practical platform for high-coupling, spurious-suppressed acoustic resonators and provide a scalable route toward wideband ultrasonic resonators, filters, and related transducers.

Piezoelectric acoustic resonators are key building blocks for ultrasonic transducers, filters, sensors, and frequency-control components. Across these applications, practical devices increasingly require the simultaneous realization of high electromechanical coupling, high linearity, low temperature drift, and spectrally stable single-mode responses. Conventional thin-film bulk acoustic resonators (FBARs) based on AlN and AlScN offer clear advantages in process maturity and manufacturability at higher frequencies, but their effective electromechanical coupling coefficient ($k_t^2$) still limits the achievable bandwidth. By contrast, strongly piezoelectric single crystals such as LiNbO3 and LiTaO$_3$ provide significantly larger piezoelectric coefficients and support a richer variety of acoustic modes—including Rayleigh waves, Love waves, Sezawa waves, Lamb waves, and higher-order thickness-shear modes—enabled by orientation engineering. These materials are therefore regarded as promising platforms for next-generation ultrasonic and acoustic components where large bandwidth, high power efficiency, and stable modal behavior are simultaneously required. [1,2]

Among these single-crystal piezoelectrics, LiTaO$_3$ offers several engineering advantages over LiNbO$_3$. Commonly used orientations of LiTaO$_3$ exhibit a smaller absolute temperature coefficient of frequency (TCF), yielding improved resonance-frequency stability; its lower intrinsic mechanical loss supports higher quality factors; and under large-signal drive it typically shows higher power tolerance and reduced nonlinear distortion. Enabled by a mature SAW-filter process flow and an established industrial supply chain, LiTaO$_3$ is being pursued for high-stability communication filters, medium- to high-power acoustic drivers, and acoustic imaging/stimulation applications. [3,4] For these use cases, a central challenge is to further increase $k_t^2$ while preserving power handling and linearity, so that wider usable bandwidth can be achieved under practical constraints on device footprint and stability.

Efforts to enhance the electromechanical coupling of LiTaO$_3$ have, over the past several decades, largely followed a "mode selection plus orientation engineering" strategy. Early work (from the late 1970s through the 1990s) focused primarily on SH-SAW modes and S$_0$-type surface/plate-wave modes. In this regime, the achievable $k_t^2$ was still modest, with typical values on the order of 3.9%–4.7%. [5-7] In the early 2000s, the adoption of zero-order shear modes (SH$_0$) marked the first major jump in coupling, raising the $k_t^2$ to roughly 13%. [8] Subsequently, around 2014–2015, further gains were achieved by exploiting specific LiTaO$_3$ orientations (e.g., 36°YX and related YX-type cuts) and optimizing quasi-shear modes (q-SH$_0$). In these configurations, the reported $k_t^2$ approached 18%, which is widely regarded as a key milestone for this material system. [9,10] Beyond that stage, development shifted toward higher operating frequencies and higher-order modes. However, these later efforts did not produce a further increase in $k_t^2$, and the reported values generally remained around that level despite continued advances in frequency scaling.[11-19] This performance still falls short of the intrinsic potential suggested by the material's piezoelectric coefficients. In summary, achieving a further increase in $k_t^2$ for LiTaO3 resonators while preserving a well-controlled single-mode spectral response remains an outstanding challenge for the field.

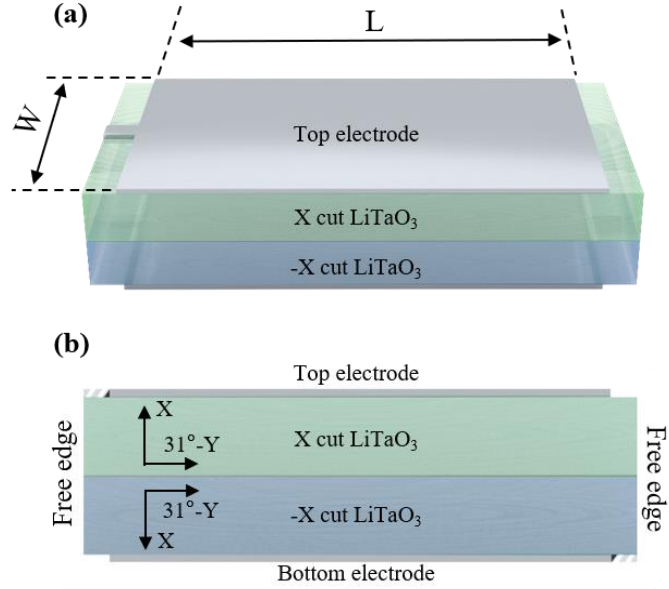

**FIG. 1.** (a) Three-dimensional diagram of the double-layer lithium tantalate structure resonator. (b) Schematic diagram of the cross-section of the resonator.

Here we report a double-layer [20-24] LiTaO3 bulk acoustic resonator that exploits symmetry-selective excitation of the second-order thickness-shear (SH2) mode. The device consists of two 31° Y-oriented LiTaO3 films that are rotation-bonded in opposite polarization (+X/−X), with Al electrodes on the top and bottom surfaces to drive the SH2 mode under a longitudinal electric field (Fig. 1). In this configuration, the resonator exhibits an experimentally measured effective electromechanical coupling coefficient of $k_t^2$ = 25.7%. Experimental characterization shows a response dominated by the target SH2 mode, with only weak parasitic features in the band of interest. Finite-element modeling further predicts a temperature coefficient of frequency (TCF) as low as approximately −24.94 ppm/°C. These results identify complementary-polarity double-layer LiTaO3 as a scalable platform for high-coupling, symmetry-engineered acoustic resonators, with practical potential for ultrasonic resonators, filters, and transducers.

In acoustic resonators and filters, the $k_t^2$ quantifies the electro-acoustic conversion efficiency. It reflects the combined influence of material constants, mode shape, electrode configuration, and boundary conditions, and is typically extracted from the separation between the parallel resonance $f_p$ and eries resonance $f_s$, as $k_t^2 = \frac{\pi^2}{8} \cdot \frac{f_p^2 - f_s^2}{f_s^2}$ .[25] Its value is determined by multiple coupled factors. At the most basic level, it depends on the intrinsic material properties. Different acoustic modes modify the effective elastic modulus, piezoelectric tensor, and dielectric permittivity of the device, and therefore change the coupling behavior. Crystal orientation and acoustic propagation direction act as key design degrees of freedom: by rotating the crystal relative to the device axes, the piezoelectric tensor is reprojected into the device frame, which can strongly enhance or suppress $k_t^2$. In addition, the spatial distribution and utilization of the driving electric field are critical. Practical transducers consist of an active region, where the field is applied, and a passive load region without field; the areal ratio between these regions directly influences the net coupling. Finally, the elastic, dielectric, and piezoelectric constants are themselves temperature dependent, so both $k_t^2$ and the resonance frequency drift with temperature. As a result, thermal stability (e.g., compensated cuts, temperature-compensated SAW structures, or composite stacks) must be co-designed together with coupling.

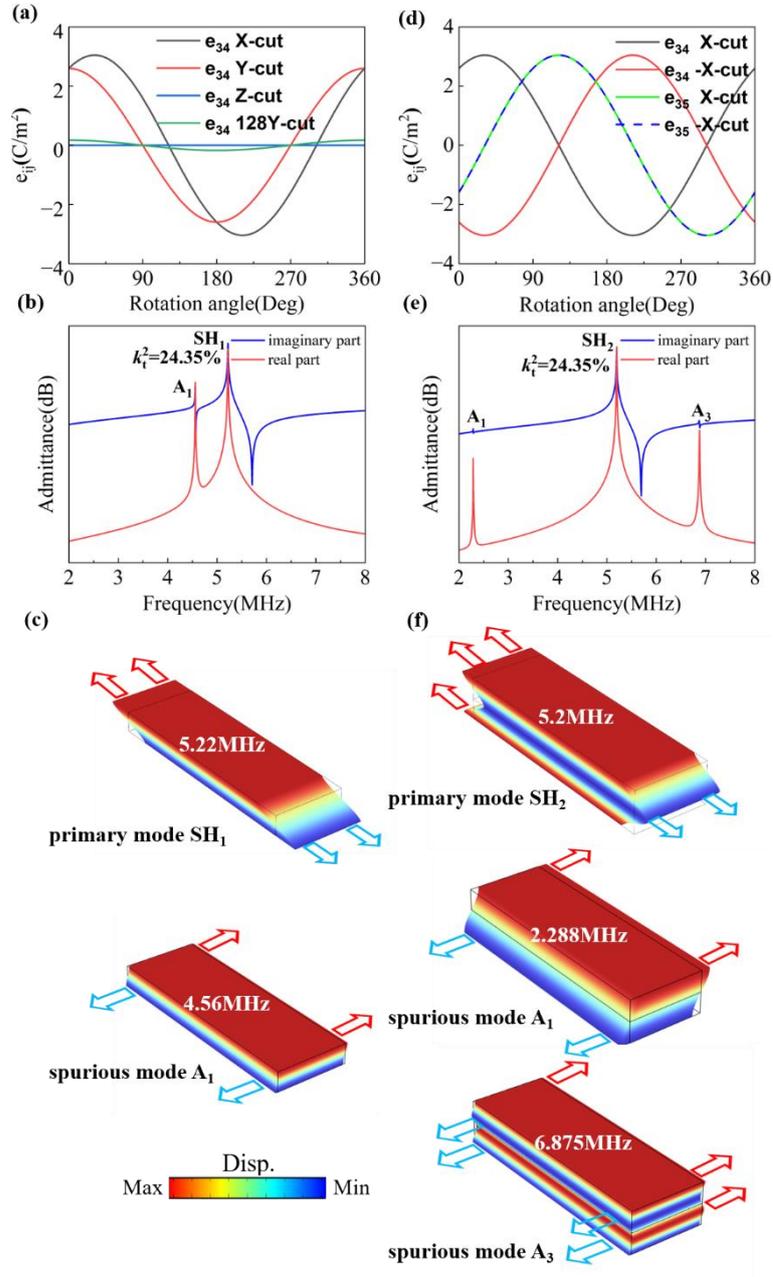

**FIG. 2.** (a) Piezoelectric coefficient $e_{34}$ of the LiTaO$_3$ at different rotation angles for four cuts. (b) Admittance spectrum of the single-layer resonator. (c) displacement-field distributions of the primary and spurious modes. (d) Piezoelectric coefficients $e_{34}$ and $e_{35}$ at different rotation angles for X and X cut LiTaO$_3$. (e) Admittance spectrum of the double-layer resonator. (f) displacement-field distributions of the primary and spurious modes.

The shear mode is electrically excited by applying an electric field along the thickness direction (X direction) of the device. For the X-cut orientation, a 31° rotated tensor projection produces the strongest effective $e_{34}$ component, making it the primary excitation pathway for this cut (Fig. 2a). In a single-layer LiTaO$_3$ resonator, the main SH$_1$ mode appears at 5.22 MHz, but a spurious A$_1$ mode driven by e$_{35}$ is also observed at 4.56 MHz. Because this A$_1$ mode is excited by a similar field direction and lies close in frequency, it is difficult to suppress in a single-layer stack (Figs. 2 b, c). To mitigate this spurious response, we adopt a double-layer LiTaO$_3$ structure formed by rotation-bonding two +X/−X oriented films. In this bilayer, different modes couple very differently (Fig. 2d). For the SH$_2$ mode, coupling is preserved: both e$_{34}$ and the electric field are antisymmetric with respect to the interface at the middle of the double-layer plate, so the drive adds constructively. The displacement profile evolves from antisymmetric (single-layer SH$_1$) to symmetric (bilayer SH$_2$), and doubling the total thickness (t→2t) does not change the resonance condition ($f_a = \frac{N\nu}{a}$, where $f_a$ is the resonant frequency, N is the mode order, ν is the acoustic phase velocity, and a denotes half of the film thickness). By contrast, for the A$_1$ mode, e$_{35}$ is symmetric with respect to the mid-plane of the double-layer plate, whereas the electric field is antisymmetric across that interface; as a result, their electromechanical drive tends to cancel. At the same time, the doubled thickness shifts the A$_1$/A$_3$ resonance condition, pushing these flexural modes away in frequency. As a result, the desired SH$_2$ mode and the unwanted A-type modes become spectrally separated [26, 27], At the same time, the A$_1$ spurious response is strongly suppressed, and the overall spectral purity of the device is improved (Figs. 2e,f). This mechanism illustrates a general strategy for spurious-mode suppression using "complementary-orientation piezoelectric films": multilayer stacks formed by +X/−X rotation-bonded LiTaO$_3$ can further enhance the purity of the target mode while suppressing undesired modes.

The resonator is realized using a bilayer LiTaO$_3$ structure consisting of a 365 μm thick +X-oriented plate and a 365 μm thick −X-oriented plate, with the LiTaO$_3$ wafers commercially supplied by NANOLN. The two wafers are rotation-bonded at room temperature with an in-plane relative rotation of 121° (starting from aligned Z axes), forming a complementary-orientation configuration (Fig. 3a). During the bonding process, the relative angle between the two plates is accurately controlled by rotating and aligning the wafer edges to achieve the desired crystallographic orientation. After bonding, a multi-step annealing process with a gradually increased temperature profile is applied to enhance the interfacial bonding strength. This stepped thermal treatment helps relieve internal stress arising from the anisotropic thermal expansion of LiTaO$_3$ along different crystallographic directions, thereby improving bonding quality and structural stability. Finally, the bonded wafer is diced into individual samples according to the designed device dimensions.

Symmetric 200 nm Al electrodes are deposited on the top and bottom surfaces to provide piezoelectric drive of the SH mode; the fabricated device and its layout are shown in (Fig. 3b). Finite-element analysis indicates that the dominant mode is SH$_2$ at 5.20 MHz, with $k_t^2$=24.35%, and a flat admittance spectrum without significant spurious responses in the band of interest (Fig. 3c). Electrical characterization was performed by aluminum wire-bonding the device and measuring the admittance ($Y_{11}$) using a vector network analyzer (Keysight E5063A). A strong SH$_2$ resonance peak is observed at 5.24 MHz, with no pronounced additional spurious peaks. The measured results show good agreement with the simulation, yielding an effective electromechanical coupling coefficient of $k_t^2$=25.70% (Fig.3c). The quality factors extracted from the measurement are 348 at the series resonance frequency ($f_s$) and $Q_p$= 601 at the parallel resonance frequency ($f_p$), indicating moderate acoustic loss and good energy confinement of the device. These results further confirm the effectiveness of the proposed structure in achieving high coupling together with controlled spectral characteristics.

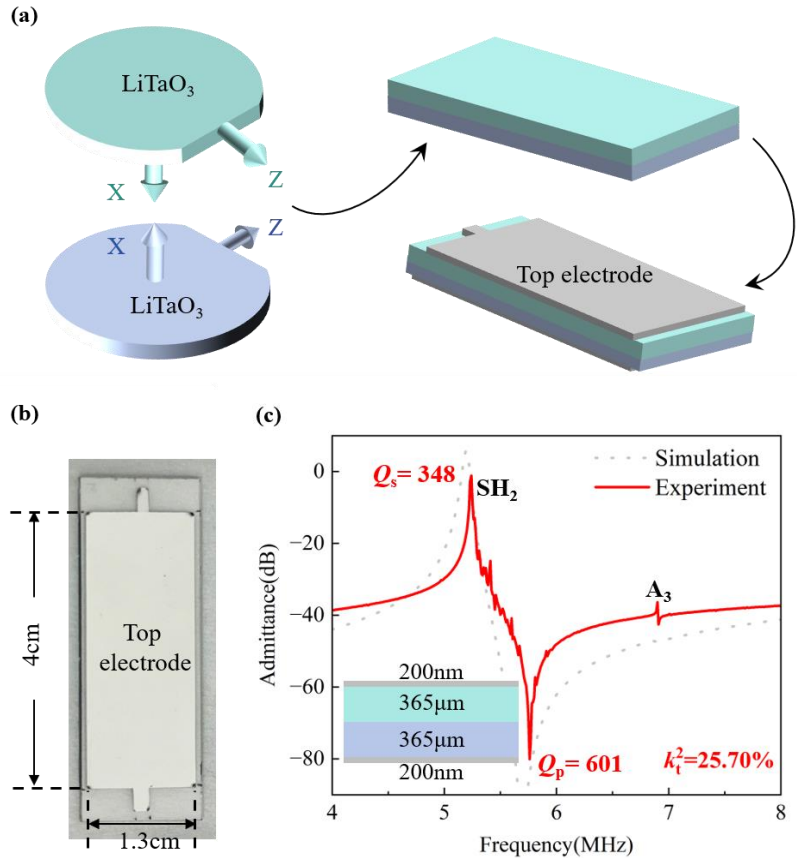

**FIG. 3.** (a) Fabrication process schematic of the device. (b) Device micrograph and structural schematic. (c) Simulated and Measured admittance spectrum.

Table I compares representative LiTaO3 resonators in terms of material configuration, operating mode, and effective electromechanical coupling. Our bilayer SH2 device exhibits the highest $k_t^2$, which is about 45% higher than the previously reported highest LiTaO3 resonator.

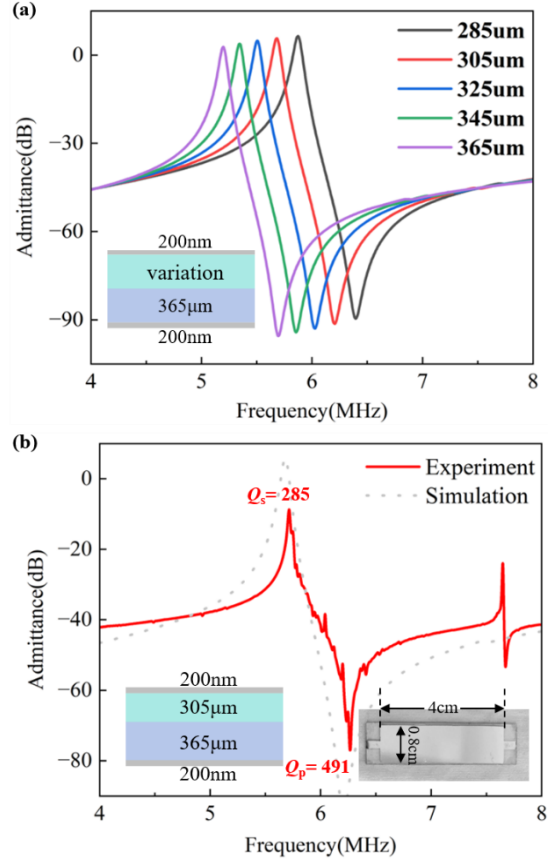

**FIG. 4.** (a) Effect of varying the single-layer film thickness. (b) Simulated and measured admittance spectra for a stack thickness of (305+365) μm.

The resonator also offers substantial design tunability and provides room for further optimization. By adjusting the in-plane rotation angle between the LiTaO3 layers ($\theta$), the film thickness ($t$), and the electrode thickness ($t_e$), the resonance frequency and effective coupling can be co-optimized for different application/design targets. Fig. 4a shows simulations in which only the single-layer thickness is varied: the resonance frequency decreases monotonically with increasing $t$, while $k_t^2$ remains high and no additional spurious modes appear in the examined range. Experimental measurements on a device with $t = 305$ μm (Fig. 4b) are consistent with the simulation: the admittance spectrum shows no new spurious responses, and the device maintains $k_t^2=24.87\%$. The extracted quality factors are $Q_s = 285$ at $f_s$ and $Q_p = 491$ at the $f_p$.

TABLE I. Performance comparison of state-of-the-art large $k^2$ piezoelectric LiTaO$_3$ resonators.

| Ref. | Cut | mode | $k_t^2$(%) |
|---|---|---|---|
| 17(2023) | 36°YX | S$_3$ | 0.72 |
| 15(2021) | 42°YX | SH$_4$ | 1.38 |
| 19(2024) | 42°YX | A$_1$ | 2.47 |
| 18(2023) | 42°YX | LL-SAW | 4.61 |
| 4(2022) | 42°YX | SAW | 9.2 |
| 13(2019) | 50°YX | SH-SAW | 9.73 |
| 14(2021) | 63°YX | SH$_1$ | 12.74 |
| 16(2022) | 42°YX | SH-SAW | 13.44 |
| 12(2019) | 42°YX | A$_1$ | 17.65 |
| **Ours** | **±31°YX** | **SH$_2$** | **25.70** |

Similarly, variations in other structural parameters produce consistent trends. (i) When only the relative in-plane rotation angle between the two LiTaO$_3$ layers is adjusted within a small range, both the operating frequency and $k_t^2$ remain essentially unchanged, as shown in Fig. 5a. (ii) When the top/bottom Al electrodes are thinned from 200 nm to 20 nm and the LiTaO$_3$ film thickness is simultaneously reduced from 365 μm to 365 nm, the resonance frequency shifts above 5 GHz while $k_t^2$ remains as high as 24.90% (Fig. 5b). Importantly, in all cases these parameter variations do not introduce pronounced additional parasitic features within the main operating band: the admittance spectra remain smooth and are dominated by the target mode, indicating good tolerance to fabrication perturbations. The difference between the MHz experimental results and the GHz simulations arises from geometric scaling. Since the resonance frequency of thickness-shear modes is inversely proportional to the structural thickness, proportional dimensional scaling enables frequency upshifting without altering the underlying physical mechanism, as also demonstrated in Ref [28].

Using reported temperature coefficients of the elastic constants, piezoelectric coefficients, and dielectric permittivity of LiTaO$_3$, [29-31] we simulated the temperature coefficient of frequency (TCF) of the bilayer LiTaO$_3$ resonator. A linear fit of the simulated resonance shift versus temperature (Fig. 5c) gives TCF ≈ −52.26 ppm/°C, consistent with typical values for uncompensated LiTaO$_3$ devices. Temperature compensation schemes commonly introduce SiO$_2$ as a stress- and velocity-compensation layer, and the same approach can be applied here. When 360 μm SiO$_2$ layers are added symmetrically to the top and bottom surfaces of the bilayer LiTaO$_3$ stack, the simulated TCF improves to −24.94 ppm/°C (Fig. 5d). Importantly, this modification does not introduce pronounced additional parasitic features within the main operating band, and the response remains dominated by the primary mode. Although the compensated structure shifts the dominant resonance from SH$_2$ to SH$_4$, the calculated $k_t^2$ remains as high as 18.97%, which is still substantial for LiTaO$_3$ resonators.

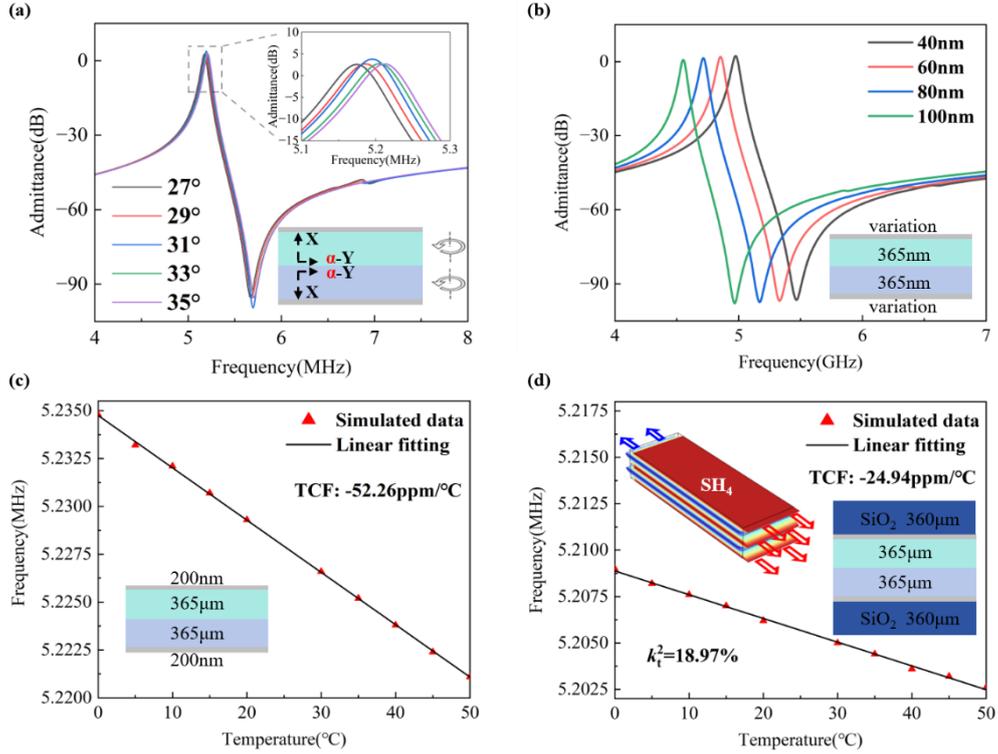

**FIG. 5.** (a) Effect of varying the relative rotation angle between the top and bottom films. (b) For a device scaled to 5 GHz: effect of electrode thickness on admittance. (c) Admittance response of a double-layer LiTaO$_3$ resonator at different temperatures. (d) Admittance response at different temperatures after adding SiO$_2$ temperature-compensation layers on both the top and bottom surfaces. SH$_2$ shift to SH$_4$ but remaining a relatively high $k_t^2$.

In conclusion, we have presented and validated a double-layer LiTaO$_3$ bulk acoustic resonator operating in the second-order thickness-shear (SH$_2$) mode. The device is formed by rotation-bonding two LiTaO$_3$ plates in a complementary +X/−X orientation and exploits symmetry matching between the piezoelectric tensor and the target mode to differentiate the SH$_2$ resonance from the parasitic A$_1$ branch. As a result, the resonator exhibits a response dominated by the target SH$_2$ mode in the band of interest, together with an ultrahigh measured effective electromechanical coupling coefficient of $k_t^2 =$ 25.7%. The platform is tunable and robust: by scaling the LiTaO$_3$ thickness and electrode thickness, the operating frequency can be extended beyond 5 GHz while maintaining weak spurious response and ≈25% $k_t^2$; variations in the relative in-plane rotation angle have only minor effects on frequency and coupling. With thin SiO$_2$ compensation layers, finite-element modeling that incorporates temperature-dependent material parameters predicts a temperature coefficient of frequency of about −24.94 ppm/°C without introducing pronounced additional parasitic features.

Although the present experimental demonstration is in the MHz regime, the underlying mechanism does not rely on that absolute frequency scale. The key ingredients—complementary-polarity bonding, longitudinally excited thickness-shear operation, and symmetry-selective mode excitation—are inherently frequency-scalable. Taken together, these results identify complementary-polarity double-layer LiTaO$_3$ as a scalable platform for high-coupling and mode-controlled acoustic resonance, with broad potential for ultrasonic resonators, filters, transducers, and other piezoelectric acoustic devices requiring strong coupling and controlled modal behavior.